\documentclass[twocolumn,showpacs,preprintnumbers,amsmath,amssymb]{revtex4}
\usepackage{dcolumn}
\usepackage{bm}
\usepackage{graphicx}
\usepackage{amsmath}
\usepackage{subfigure}
\usepackage{changes}
\usepackage{graphicx}
\usepackage{amsmath}
\usepackage{hyperref}

\begin{document}

\title{Wide band magnetometry based on color centers with transverse zero-field splitting}
\author{Xiayan Ding, Lu Chen, and Qiong Chen\footnote{E-mail:qchen@hunnu.edu.cn}}
\affiliation{Key Laboratory of Low-Dimensional Quantum Structures and Quantum Control of Ministry of Education, \\Synergetic Innovation Center for Quantum Effects and Applications, Xiangjiang Laboratory, and Department of Physics, \\Institute of Interdisciplinary Studies, Hunan Normal University, Changsha, Hunan 410081, China}

\begin{abstract}
Solid-state color centers embedded in diamond, silicon carbide (SiC) and other host-matrices offer a promising platform for nanoscale quantum sensing. Sometimes a relatively strong transverse zero-field splitting (ZFS) is introduced in a color center due to its local strain and low symmetry of the host structure. While clock transitions induced by transverse zero-field splitting (ZFS) serve as a method to prolong the coherence time, the Zeeman sub-levels of solid-state color centers become insensitive to first-order magnetic field signals, thereby limiting their utility as magnetometers. In this work, we address this challenge and achieve wide band AC field detection spanning from hundreds of kHz to hundreds of MHz by utilizing a combination of orthogonal microwaves and phase modulation at different frequencies. Our control method effectively suppresses the system noise and the amplitude fluctuation of the driving field, which extends the coherence time of the quantum system.
\end{abstract}
\maketitle

\section{Introduction}

Solid-state spin color centers, such as nitrogen-vacancy (NV) centers \cite{doherty2013nitrogen,2017Quantum} in diamond and defects in silicon carbide (SiC) \cite{castelletto2020silicon,luo2023fabrication}, have emerged as crucial platforms for quantum computing, quantum networks, and quantum sensing. Among these candidates, solid-state spin sensors represented by NV centers have demonstrated remarkable achievements in magnetic field \cite{wang2022picotesla}, electric field \cite{bian2021nanoscale}, and temperature \cite{CHOE20181066} measurements, owing to their excellent room temperature quantum properties, optical operation feasibility, and diamond host stability. Recently, molecular color centers have garnered significant attention due to their atomic-scale extensibility, chemical tunability and biocompatibility \cite{chiesa2023chirality,mena2024room,gaita2019molecular,fursina2023toward,wasielewski2020exploiting,yu2021molecular,kopp2024luminescent}. A principal advantage of molecular color centers lies in their capacity for customization, facilitating direct attachment to the surfaces or within target substrates, thereby minimizing sensing distance between the sensors and targets \cite{mullin2023quantum}. Despite their advantages, the performance of these solid-state color centers is typically constrained by magnetic noise from surrounding nuclear spin baths \cite{hanson2008coherent}, which will shorten the coherence time of the quantum systems. 

Fortunately, transverse zero-field splitting (ZFS) can induce clock transitions near zero magnetic field, effectively rendering the transition frequencies between spin states insensitive to magnetic field variations \cite{miao2020universal,liu2021quantum,miao2019electrically}. Additionally, sometimes transverse ZFS is unavoidable for molecular color centers because of local deformation or symmetry of the host structure \cite{bayliss2022enhancing,awadallah2023spin,zhu2021theoretical}. While the clock transition serves to suppress magnetic noise and extend the coherence time, it also becomes insensitive to weak magnetic signals, complicating the detection of alternating current (AC) fields.
Furthermore, surface electric noise \cite{kim2015decoherence} and second-order magnetic noise induced by nuclear spin baths further accelerate the decoherence of the system \cite{jamonneau2016competition,yang2016quantum}, presenting further challenges that necessitate innovative solutions.

To address these challenges, our method employs microwave (MW) drives with distinct frequencies along orthogonal spatial axes along the $x$ and $y$ direction, in contrast to conventional linear MW driving \cite{saijo2018ac,stark2018clock} aligned perpendicular to the solid-state spin quantization axis ($z$-axis).  By leveraging solid-state color centers with transverse zero-field splitting (ZFS), our approach achieves broadband magnetometry, thereby enabling a detection range spanning from hundreds of kHz to hundreds of MHz. The utilization of phase-modulated MW driving \cite{farfurnik2017experimental,cao2020protecting} serves to reduce the amplitude fluctuations in the control field, thereby enhancing the coherence of the quantum system.

The structure of this paper is as follows. We begin by presenting a system model that emphasizes clock transitions induced by transverse zero-field splitting (ZFS) in solid-state color centers. Next, we validate our control scheme by comparing its robustness to system noise with traditional linear microwave (MW) driving. As amplitude fluctuations in the driving field further reduce coherence time, we incorporate phase modulation as a countermeasure. Finally, we apply our scheme to AC field detection, and numerical simulations demonstrate an expanded detection range for AC fields and improved coherence time in solid-state spins compared to molecular clock sensors employing continuous radio-frequency driving.

\section{the control methods}
\subsection{The model}
We start by considering a solid-state spin with a spin-triplet ground state ($S=1$) in the case of a relatively large transverse zero-field splitting (ZFS). The magnetic field $B$ is applied along the spin axis ($z$-axis).
The Hamiltonian for the solid-state spin in the absence of a microwave (MW) driving field is then given by (with $\hbar=1$)
\begin{align}
H=DS_{z}^{2}+[E_{x}+\delta E(t)](S_{x}^{2}-S_{y}^{2})+[\gamma_{e}B_{z}+\delta_{B_{z}}(t)]S_{z}.
\end{align}
Here, $D$ represents the axial ZFS between the $M_{s}=0$ and $M_{s}=\pm1$ spin sublevels, $E_{x}$ denotes the transverse ZFS, and $\gamma_{e}$ is the gyromagnetic ratio of the electron. $\delta E(t)$ is the strain/electric field noise and $\delta_{B_{z}}(t)$ represents the magnetic noise. In the following analysis, we denote them by $\delta E$ and $\delta_{B_{z}}$ for simplicity. The states $|M_{s}\rangle=|0\rangle, |\pm1\rangle$ correspond to the eigenstates of the solid-state spin operator $S_{z}$ and $\mathbf{S}=(S_{x},S_{y},S_{z})$ are the dimensionless spin-1 operators for the solid-state spin. Specifically, $S_{z}$=$|+1\rangle\langle +1|-|-1\rangle\langle -1|$ and $S_{x}$=$(+1/\sqrt{2})(|+1\rangle\langle 0|+|-1\rangle\langle 0|+h.c.)$. The eigenenergies and eigenstates of the Hamiltonian $H$ are given by
\begin{align}
\omega_{+1}&=D+\sqrt{(E_{x}+\delta E)^{2}+(\gamma_{e}B_{z}+\delta_{B_{z}})^{2}},\\
\omega_{-1}&=D-\sqrt{(E_{x}+\delta E)^{2}+(\gamma_{e}B_{z}+\delta_{B_{z}})^{2}},\\
\omega_{0}&=0,
\end{align}
and
\begin{align}
|\psi_{+}\rangle&=\cos(\frac{\theta}{2})|+1\rangle+\sin(\frac{\theta}{2})|-1\rangle, \\
|\psi_{-}\rangle&=-\sin(\frac{\theta}{2})|+1\rangle+\cos(\frac{\theta}{2})|-1\rangle,\\
|\psi_{0}\rangle&=|0\rangle.
\end{align}
The rotation angle $\theta$ is given by $\cos\theta=(\gamma_{e}B_{z}+\delta_{B_{z}})/\sqrt{(E_{x}+\delta E)^{2}+(\gamma_{e}B_{z}+\delta_{B_{z}})^{2}}$.

When a strong magnetic field is applied, i.e., $\gamma_{e}B_{z}\gg E_{x}$, corresponding to $\theta\approx0$, the decoherence of the solid-state spin is primarily dominated by magnetic field noise $\delta_{B_{z}}$ \cite{jamonneau2016competition,PhysRevLett.112.147602,RevModPhys.92.015004}, arising from the coupling between the magnetic field and neighboring nuclear spins. For example, in a noisy nuclear spins molecular system, such as Cr(IV)-based Cr(IV)(aryl)$_{4}$, $\delta_{B_{z}}$ is typically on the order of MHz \cite{janicka2022computational}. It could also be feasible to employ 99.998\% $^{12}$C purity to eliminate the nuclear spin noise bath of NV center \cite{alsid2023solid}.

Here, we consider the general case of $E_{x}\gg\gamma_{e}B_{z}$, corresponding to $\theta\approx\pi/2$, the energy levels of the system exhibit the states $|\mu_{\pm}\rangle_{c}=(1/\sqrt{2})(|-1\rangle\pm|+1\rangle)$ near magnetic field zero point. The Hamiltonian of the system can be approximated as
\begin{align}
H_{0}\approx DS_{z}^{2}+(E_{x}+\delta E)(S_{x}^{2}-S_{y}^{2}).
\end{align}
The effective Hamiltonian of the two-level system is expressed as
\begin{align}
H_{\mu}^{c}=2(E_{x}+\delta E)\sigma_{z}^{c},
\end{align}
where $\sigma_{z}^{c}=(1/2)(|\mu_+\rangle_{c}\langle \mu_+|-|\mu_-\rangle_{c}\langle \mu_-|)$.
The eigenstates of $H_{0}$ are $|0\rangle$, $|\mu_{-}\rangle_{c}$, and $|\mu_{+}\rangle_{c}$, with corresponding eigenvalues 0, $D-E_{x}$, and $D+E_{x}$, respectively. The energy difference between $|\mu_{+}\rangle_{c}$ and $|\mu_{-}\rangle_{c}$ is $\omega_{+}-\omega_{-}=2E_{x}$. In this regime, ``clock transitions'' make spin transitions insensitive to first-order magnetic field noise. When the system satisfies $E_{x}\gg\gamma_{e}B_{z}, \delta_{B_{z}}$, magnetic noise $\delta_{B_{z}}$ is suppressed in the $|\mu_{\pm}\rangle_{c}$ states, approximately $\sim\delta_{B_{z}}^{2}/2E_{x}$, which reduces the noise to the kHz range (as shown in Fig. 1). As a result, the solid-state spin is protected from first-order magnetic field fluctuations and it is reasonable to assume that $\delta E$ (hereafter referred to as system noise), predominantly comprises strain/electric field noise and second-order magnetic fluctuations \cite{mena2024room,JiangChen2024}.

\begin{figure}[h]
  \centering
  \includegraphics[scale=0.5]{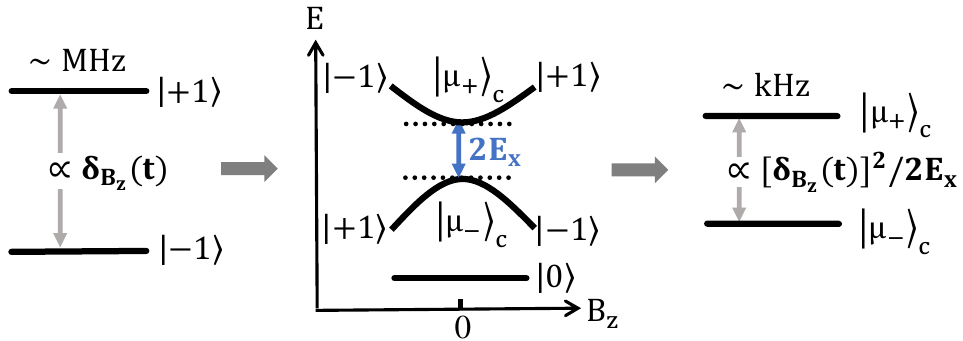}
  \caption{ The energy level structure of a color center as a function of $B_z$ (\textit{middle}). When $|\gamma_{e}B_{z}|\gg E_{x}$ (\textit{left}), the solid-state color center basis is given by $|\pm1\rangle$. When $E_{x}\gg\gamma_{e}B_{z}$  (\textit{right}), and we have states $|\mu_{\pm}\rangle_{c}$ via clock transitions, effectively suppressing the magnetic field noise. Here the state $|\psi_{0}\rangle$ is not included.}\label{tu1}
\end{figure}
\vspace{-1em}

\subsection{Linear MW driving}
Traditional quantum control methods primarily involve applying linearly polarized microwave (MW) fields to drive transitions between the energy levels of color center spins, enabling effective control of the quantum system \cite{kong2018nanoscale}.
Using a resonant MW with the Hamiltonian $H_{c1}=2\Omega\cos(\omega t)S_{x}$, where $\Omega$ represents the MW driving amplitude (Rabi frequency) and $\omega$ is the MW frequency, the system is transformed into a rotating frame with respect to $H_{0}^{'}=DS_{z}^{2}+E_{x}(S_{x}^{2}-S_{y}^{2})$. This results in the system Hamiltonian becoming $H_{1I}=\Omega S_{x}$, assuming the resonance condition $\omega=D+E_{x}$. To derive this, the high-frequency oscillation term is neglected under the rotating wave approximation (RWA), assuming $\Omega\ll2\omega$.\par

When $E_{x}\gg\gamma_{e}B_{z}$, the decoherence of the solid-state color center under clock transitions is predominantly governed by $\delta E$. To elucidate the performance of linear MW driving against this decoherence source $\delta E$, we consider the Hamiltonian containing $\delta E$ as
\begin{align}
H^{'}_{1}=DS_{z}^{2}+(E_{x}+\delta E)(S_{x}^{2}-S_{y}^{2})+2\Omega\cos(\omega t)S_{x}.
\end{align}
By moving to the rotating frame with respect to $H_{0}^{'}$ and applying the resonance condition $\omega=D+E_{x}$ under the RWA, where $\Omega\ll2\omega$, we can rewrite this Hamiltonian as
\begin{align}
H^{'}_{1I}=\begin{pmatrix}
0 &\frac{\Omega}{\sqrt{2}} & \delta E \\
\frac{\Omega}{\sqrt{2}} & 0 & \frac{\Omega}{\sqrt{2}} \\
\delta E &\frac{\Omega}{\sqrt{2}}& 0
\end{pmatrix},
\end{align}
 in the basis $\{|+1\rangle,|0\rangle,|-1\rangle\}$. The eigenvalues of the matrix are as follows
\begin{align}
\omega_{\mu_{+}}&=\frac{1}{2}[\delta E-\sqrt{4\Omega^{2}+(\delta E)^{2}}],\\
\omega_{\mu_{-}}&=\frac{1}{2}[\delta E+\sqrt{4\Omega^{2}+(\delta E)^{2}}],\\
\omega_{\mu}&=-\delta E,
\end{align}
and the eigenstates of the matrix are
\begin{align}
|\mu_{+}\rangle_{l}&=\frac{1}{\sqrt{2}}\cos\eta(|+1\rangle+|-1\rangle)+\sin\eta|0\rangle, \\
|\mu_{-}\rangle_{l}&=\frac{1}{\sqrt{2}}\sin\eta(|+1\rangle+|-1\rangle)-\cos\eta|0\rangle,\\
|\mu\rangle_{l}&=-\frac{1}{\sqrt{2}}|+1\rangle+\frac{1}{\sqrt{2}}|-1\rangle,
\end{align}
with rotation angle relation
\begin{align}
\sin\eta&=\frac{r_{+}}{\sqrt{2+r_{+}^{2}}},\cos\eta=\frac{r_{-}}{\sqrt{2+r_{-}^{2}}}
\end{align}
and
\begin{align}
r_{\pm}=\frac{\sqrt{4\Omega^{2}+(\delta E)^{2}}\mp\delta E}{\sqrt{2}\Omega}.
\end{align}
In the new basis of interaction picture, the energy gap of the two-level system of $\{|\mu\rangle_{l},|\mu_{-}\rangle_{l}\}$ is given by
\begin{align}
\Delta\omega_{l}&=-\frac{3}{2}\delta E-\frac{1}{2}\sqrt{4\Omega^{2}+(\delta E)^{2}}\\
&\approx-\frac{3}{2}\delta E-\Omega-\frac{(\delta E)^{2}}{8\Omega}\nonumber,
\end{align}
and the effective Hamiltonian becomes
\begin{align}
H_{\mu}^{l}=\Delta\omega_{l}\sigma_{z}^{l},
\end{align}
where $\sigma_{z}^{l}=(1/2)(|\mu\rangle_{l}\langle \mu|-|\mu_-\rangle_{l}\langle \mu_-|)$.
The energy gap $\Delta\omega_{l}$ (see Fig. 2) exhibits nearly linear dependence on the system noise $\delta E$, indicating that the system is not robust to $\delta E$ under linear MW driving. \par

\begin{figure}[h]
  \centering
  \includegraphics[scale=0.5]{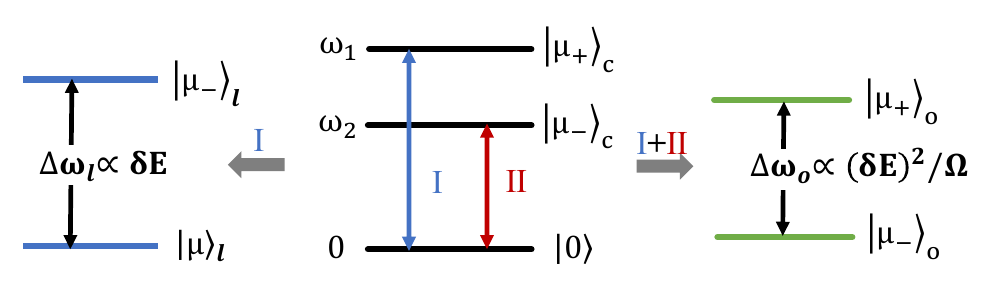}
  \caption{Different MW engineering schemes of the solid state spin when there is a relatively large transverse ZFS. The energy level structure of a color center is shown when $E_{x}\gg\gamma_{e}B_{z}$ (\textit{middle}). The effective two-level system with traditional linear MW drive of states $|0\rangle$ and $|\mu_{+}\rangle_{c}$ is not robust to $\delta E$ (\textit{left}). Our control scheme, by driving both $|0\rangle$ and $|\mu_{+}\rangle_{c}$ and $|0\rangle$ and $|\mu_{-}\rangle_{c}$ states simultaneously, can show a robustness against $\delta E$ (\textit{right}).}\label{tu1}
\end{figure}

\subsection{Our control method}
 In order to protect against decoherence due to $\delta E$, we propose an innovative method which employs MW drives with
distinct frequencies along the $x$- and $y$-axis, respectively. The Hamiltonian of solid-state spin with MW control is written as
\begin{align}
H_{2}&=DS_{z}^{2}+(E_{x}+\delta E)(S_{x}^{2}-S_{y}^{2})\\&+\sqrt{2}\Omega\cos(\omega_{1} t)S_{x}+\sqrt{2}\Omega\sin(\omega_{2} t)S_{y}.\nonumber
\end{align}

As shown in Fig. 2, we apply a MW field with frequencies $\omega_{1}$ along the $x$-axis to drive the transition $|0\rangle\leftrightarrow|\mu_{+}\rangle_{c}$, while simultaneously employing a MW field with frequencies $\omega_{2}$ along the $y$-axis to drive $|0\rangle\leftrightarrow|\mu_{-}\rangle_{c}$. In the rotating frame with respect to $H_{0}^{'}=DS_{z}^{2}+E_{x}(S_{x}^{2}-S_{y}^{2})$, we select $\omega_{1}=D+E_{x}$ and $\omega_{2}=D-E_{x}$ to resonant with the $|0\rangle\leftrightarrow|\mu_{+}\rangle_{c}$ and $|0\rangle\leftrightarrow|\mu_{-}\rangle_{c}$ transitions, respectively. Under continuous MW selective drive transitions, the above Hamiltonian can be represented as a matrix
\begin{equation}
H_{2I}^{'}=\begin{pmatrix}
0 &\Omega & \delta E \\
\Omega & 0 & 0 \\
\delta E &0& 0
\end{pmatrix},
\end{equation}
 in the basis $\{|+1\rangle,|0\rangle,|-1\rangle\}$. When we have $\delta E\ll\Omega$, after applying the RWA ($\Omega\ll2\omega_{1}, 2\omega_{2}$) and neglecting high-frequency oscillation terms, the solid-state spin is effectively treated as a qubit represented by the states $|0\rangle$ and $|+1\rangle$. Then, the Hamiltonian simplifies to
\begin{equation}
H_{2I}\approx2\sqrt{\Omega^{2}+(\delta E)^{2}}{\sigma}_{x},
\end{equation}
where $\sigma_{z}=(1/2)(|+1\rangle\langle +1|-|0\rangle\langle 0|)$, and $\sigma_{x}=(1/2)(|0\rangle\langle +1|+|+1\rangle\langle 0|)$. Here, it is easy to rotate the system by $-\pi/4$ around the $y$-axis, such that $\sigma_{x}\rightarrow\sigma_{z}^{o}$, $\sigma_{z}\rightarrow-\sigma_{x}^{o}$, and $\sigma_{y}\rightarrow\sigma_{y}^{o}$. Thus, we obtain the effective Hamiltonian
\begin{equation}
H_{\mu}^{o}=\Delta\omega_{o}\sigma_{z}^{o},
\end{equation}
where $\sigma_{z}^{o}=(1/2)(|\mu_+\rangle_{o}\langle \mu_+|-|\mu_-\rangle_{o}\langle \mu_-|)$ with $|\mu_{\pm}\rangle_{o}=(1/\sqrt{2})(|0\rangle\ \pm|+1\rangle)$. Then, the energy gap of the effective two-level system in the interaction picture is given by
\begin{align}
\Delta\omega_{o}&=2\sqrt{\Omega^{2}+(\delta E)^{2}}
\approx2\Omega+\frac{(\delta E)^{2}}{\Omega}.
\end{align}
The energy gap is demonstrated to scale as $(\delta E)^{2}/\Omega$, resulting in enhanced coherence times due to suppressed decoherence pathways in the quantum system.

\subsubsection*{phase modulation}

\begin{figure}[h]
  \centering
  \includegraphics[scale=0.5]{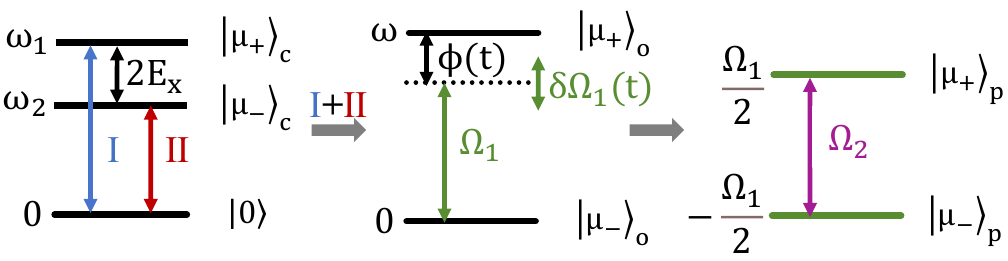}
  \caption{ Schematic of level structure by using phase modulation. The amplitude noise $\delta\Omega_{1}(t)$ is reduced by the second driving $\Omega_{2}$ in dressed states.}\label{tu1}
\end{figure}

While continuous dynamic decoupling (CDD) \cite{hirose2012continuous} has been established as an effective strategy for suppressing environmental noise \cite{genov2019mixed}, it intrinsically introduces additional noise originating from the driving field itself. Amplitude fluctuations in the control field fundamentally constrain the achievable coherence time, necessitating a hierarchical driving protocol where a progressively attenuated driving source is applied to counteract the noise contributions from the preceding order of the dressing field \cite{wang2020coherence,cai2012robust}. Here, we employ an alternative approach, where a time dependent phase is added to CDD, yielding a time-dependent
detuning \cite{cohen2017continuous}. We consider amplitude noise $\delta\Omega_{1}(t)$ in the first driving amplitude $\Omega_{1}$, leading to the following Hamiltonian
\begin{align}
H_{2}^{''}&=DS_{z}^{2}+(E_{x}+\delta E)(S_{x}^{2}-S_{y}^{2})+\\
&\sqrt{2}[\Omega_{1}\hspace{-0.2em}+\hspace{-0.2em}\delta\Omega_{1}(t)]\{\cos[\omega_{1} t\hspace{-0.2em}+\hspace{-0.2em}\phi(t)]S_{x}\hspace{-0.2em}+\hspace{-0.2em}\sin[\omega_{2} t+\phi(t)]S_{y}\}\nonumber,
\end{align}
where $\Omega_{1}$ is Rabi frequency of the continuous driving field,  $\phi(t)=2\frac{{\Omega}_{2}}{{\Omega}_{1}}\sin(2\Omega_{1}t)$ is time-dependent phase modulation, and a second driving $\Omega_{2}$ controls the phase
modulation strength (see Fig. 3). In the rotating frame with respect to $H_{0}^{'}=DS_{z}^{2}+E_{x}(S_{x}^{2}-S_{y}^{2})$ and we have $\omega_{1}=D+E_{x}$ and $\omega_{2}=D-E_{x}$. When the counter-rotating terms
are neglected, the Hamiltonian in the first rotating frame is described by
\begin{align}
H_{2I}^{''}=2[\Omega_{1}\hspace{-0.2em}+\hspace{-0.2em}\delta E^{'}\hspace{-0.2em}+\hspace{-0.2em}\delta\Omega_{1}(t)][{\cos\phi(t)\sigma}_{x}^{p}+\sin\phi(t){\sigma}_{y}^{p}],
\end{align}
where $\sigma_{z}^{p}=(1/2)(|\mu_+\rangle_{p}\langle \mu_+|-|\mu_-\rangle_{p}\langle \mu_-|)$ and $\sigma_{x}^{p}=(1/2)(|\mu_+\rangle_{p}\langle \mu_-|+|\mu_-\rangle_{p}\langle \mu_+|)$. Here we have a two-level system consisting of states $|\mu_+\rangle_{p}=|+1\rangle$ and $|\mu_-\rangle_{p}=|0\rangle$. Assuming $\delta E\ll\Omega_{1}$, the second-order effect $\delta E^{'}\approx(\delta E)^{2}/2\Omega_{1}$ is also considered.

When $\Omega_{2}\ll\Omega_{1}$, we get $\cos\phi(t)\rightarrow 1$, $\sin\phi(t)\rightarrow \phi(t)=2\frac{{\Omega}_{2}}{{\Omega}_{1}}\sin(2\Omega_{1}t)$. Under these assumptions, the Hamiltonian becomes
\begin{equation}
H_{2I^{'}}^{''}\approx2[\Omega_{1}+\delta E^{'}+\delta\Omega_{1}(t)]{\sigma}_{x}^{p}+4\Omega_{2}\sin(2\Omega_{1}t){\sigma}_{y}^{p}.
\end{equation}
\par
It is clear that phase modulation introduces a second driving that drives the dressed states on resonance with $2\Omega_{1}$. By moving to the second rotating frame with respect to $H_{2I}=2\Omega_{1}{\sigma}_{x}^{p}$, the effective Hamiltonian in the second interaction picture yields
\begin{equation}
H_{\mu}^{p}\approx2[\delta E^{'}+\delta\Omega_{1}(t)]{\sigma}_{x}^{p}+2\Omega_{2}{\sigma}_{z}^{p}.
\end{equation}
In this scheme, the time-dependent phase $\phi(t)$ suppresses the Rabi frequency noise $\delta\Omega_{1}(t)$. When both the amplitude noise $\delta\Omega_{1}(t)$ and $\delta E^{'}$ are much smaller than the second driving field $\Omega_{2}$, they can be significantly reduced, leading to increased coherence time of the solid-state spins. Moreover, there is no fluctuation in $\Omega_{2}$ because of the possible precise control of MW phase.\par

\subsection{Numerical simulation}
To demonstrate the efficiency of our proposal, we consider a color center with $E_{x}=(2\pi)\times24$ MHz in zero magnetic field and compare the robustness to noise under four different controls: (1) no driving with Eq.(9), (2) linear MW driving with Eq.(21), (3) our proposed control method with Eq.(25), and (4) our control method optimized with phase modulation with Eq.(30). Assuming the system is initialized at the state $|\psi\rangle=(1/\sqrt{2})(|\mu_{(+)}\rangle_{i}+|\mu_{-}\rangle_{i})$, where $i=c,l,o,p$ corresponds to different bases depending on the control scheme. The average value $2\langle\sigma_{x}^{i}\rangle$ (where $\sigma_{x}^{i}=(1/2)(|\mu_{(+)}\rangle_{i}\langle\mu_{-}|+|\mu_{-}\rangle_{i}\langle \mu_{(+)}|)$) reflects the robustness to noise. The results are depicted in Fig. 4. \par
  
\begin{figure}
  \centering
  \includegraphics[scale=0.5]{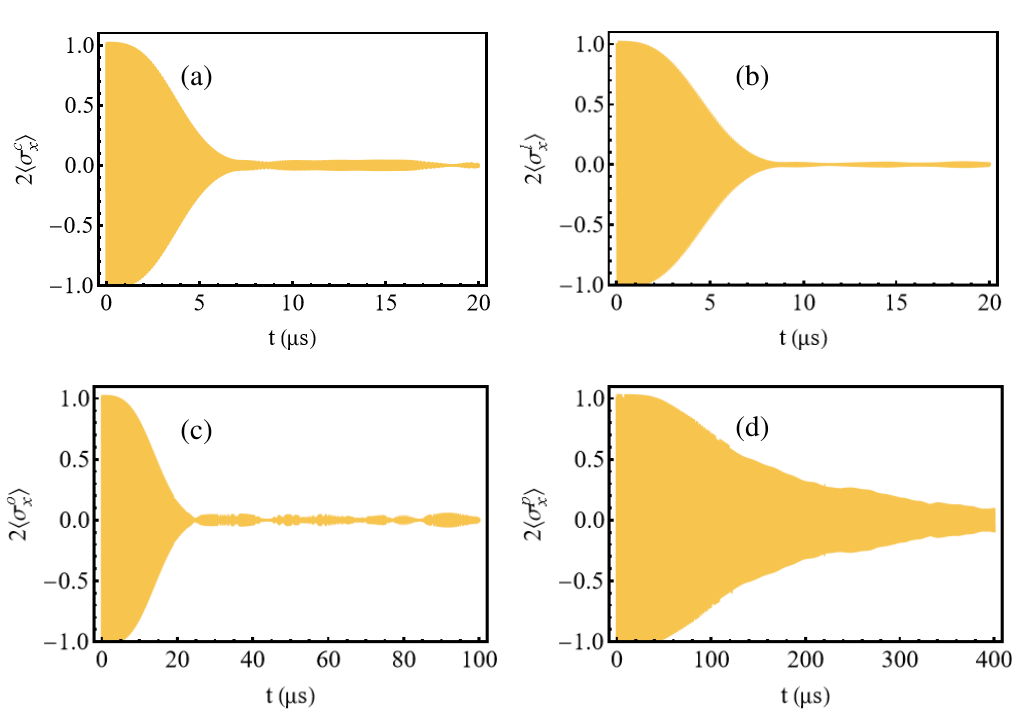}
  \caption{ Simulations under different control schemes, which is averaged over 5000 trials. All simulations use the following parameters: MW frequency $\Omega_{1}$ (or $\Omega)=(2\pi)\times 10$ MHz, $\Omega_{2}=(2\pi)\times1$ MHz and $E_{x}=(2\pi)\times24$ MHz. (a) Pure dephasing with no driving. The system is limited to the system noise $\delta E$, leading to the pure dephasing time $T_{2}^{\ast}=3$ $\mu s$. (b) Dephasing due to $\delta E$ under linear MW driving. The coherence time $T_{2}$ is approximately 4 $\mu s$. (c) Dephasing under our control method. The system noise $\delta E$ is suppressed to second-order effect and the primary limitation shifts to the amplitude noise $\delta\Omega_{1}(t)$. The relative amplitude error is $0.01$. (d) Dephasing under our control scheme optimized with phase modulation. This approach results in a coherence time over 100 $\mu s$. }\label{protocol12}
\end{figure}

For the theoretical modeling of these noise ($\delta E$, $\delta_{\Omega}$), we use an Ornstein-Uhlenbeck (OU) process \cite{de2010universal,fahrurrachman2023simulating} which is realized by an exact algorithm. For example, we model the system noise $\delta E=\delta E(t)$ by an OU process with a zero expectation value $\langle\delta E(t)\rangle=0$, the stochastic function is defined by
\begin{equation}
\delta E(t+\Delta t)=\delta E(t)e^{-\frac{\Delta t}{\tau}}+n\sqrt{\frac{c\tau}{2}(1-e^{-\frac{2\Delta t}{\tau}})},
\end{equation}
where $n$ is a unit Gaussian random number, $c$ is the diffusion constant and $\tau$ is the correlation time of the noise.
For the system noise $\delta E$, we assume a correlation time $\tau$=20 $\mu s$ and a diffusion constant $c\approx4/{T_{2}^{\ast}}^{2}\tau$, where $T_{2}^{\ast}=3$ $\mu s$ is the pure dephasing time of the system without any driving. For the amplitude noise $\delta\Omega_{1}(t)=\delta_{\Omega}\Omega_{1}$ which is modelled by an uncorrelated OU process, we set the correlation time $\tau_{\Omega}=500$ $\mu s$ and the relative amplitude error $\delta_{\Omega}=0.01$. The diffusion constant is given by $c_{\Omega}=2(\delta_{\Omega}\Omega_{1})^{2}/\tau_{\Omega}$.\par

In the absence of any driving, the system undergoes pure dephasing due to the system noise $\delta E$ resulting in a pure dephasing time $T_{2}^{\ast}$ of approximately 3 $\mu s$, as shown in Fig. 4(a). When linear MW driving is applied, as depicted in Fig. 4(b), the system remains primarily limited by the system noise $\delta E$, and there is almost no improvement in coherence time. With our proposed control method, the coherence time can be extended. But it could be limited by  amplitude error (a relatively large error $0.01$ is considered here) and approximated to be 11 $\mu s$,  as illustrated in Fig. 4(c). In this regime, the system noise is suppressed to a second-order effect $\propto(\delta E)^{2}/2\Omega_{1}$, and the primary limitation shifts to the amplitude noise $\delta\Omega_{1}(t)$. Finally, by introducing phase modulation to our control scheme, we further suppress the amplitude noise $\delta\Omega_{1}(t)$, leading to a coherence time exceeding 100 $\mu$s, as shown in Fig. 4(d). The phase modulation effectively reduces the first-order amplitude noise, and the remaining decoherence is primarily due to second-order effect of the amplitude noise and other higher-order noise sources. 


\section{ applications and discussion}
We now apply the scheme proposed in this paper to sense an AC magnetic field aligned with the solid-state spin axis. The Hamiltonian is given by
\begin{align}
H_{s}&=DS_{z}^{2}+(E_{x}+\delta E)(S_{x}^{2}-S_{y}^{2})+g\cos(\omega_{ac} t)S_{z}+\\&\sqrt{2}[\Omega_{1}\hspace{-0.2em}+\hspace{-0.2em}\delta\Omega_{1}(t)]\{\cos[\omega_{1} t\hspace{-0.2em}+\hspace{-0.2em}\phi(t)]S_{x}\hspace{-0.2em}+\hspace{-0.2em}\sin[\omega_{2} t+\phi(t)]S_{y}\}\nonumber,
\end{align}
where g represents the signal strength and $\omega_{ac}$ denotes its frequency. Going to the interaction picture with respect to $H_{0}^{'}$, satisfying $\omega_{1}=D+E_{x}$ and $\omega_{2}=D-E_{x}$, and applying the RWA ($\Omega_{1}\ll2\omega_{1},2\omega_{2}$), the total Hamiltonian of a two-level system \{$|\mu_+\rangle_{p}$,$|\mu_-\rangle_{p}$\} in the first interaction picture is described by
 \begin{align}
H_{I,s}\simeq&2[\Omega_{1}+\delta E^{'}+\delta\Omega_{1}(t)][{\cos\phi(t)\sigma}_{x}^{p}+\sin\phi(t){\sigma}_{y}^{p}]\\
&+g\cos(2E_{x} t)\cos(\omega_{ac} t){\sigma}_{z}^{p},\nonumber
\end{align}
where  $\sigma_{z}^{p}=(1/2)(|\mu_+\rangle_{p}\langle \mu_+|-|\mu_-\rangle_{p}\langle \mu_-|)$ and $\sigma_{x}^{p}=(1/2)(|\mu_+\rangle_{p}\langle \mu_-|+|\mu_-\rangle_{p}\langle \mu_+|)$. Under
the assumption $\Omega_{2}\ll\Omega_{1}$, we obtain
\begin{align}
H_{I^{'},s}=&2[\Omega_{1}+\delta E^{'}+\delta\Omega_{1}(t)]{\sigma}_{x}^{p}+4\Omega_{2}\sin(2\Omega_{1}t){\sigma}_{y}^{p}\\&+
g\cos(2E_{x} t)\cos(\omega_{ac} t){\sigma}_{z}^{p}. \nonumber
\end{align}
We then move to the second interaction picture with respect to $H_{2I}=2\Omega_{1}\sigma_{x}^{p}$ and neglect the high-frequency terms $\cos(4\Omega_{1}t)$ and $\sin(4\Omega_{1}t)$, so the Hamiltonian is given by
\begin{align}
H_{II,s}&\simeq2[\delta E^{'}+\delta\Omega_{1}(t)]{\sigma}_{x}^{p}+2\Omega_{2}{\sigma}_{z}^{p}+\\&
g\cos(2E_{x} t)\cos(\omega_{ac} t)[{\sigma}_{z}^{p}\cos(2\Omega_{1}t)\hspace{-0.2em}-\hspace{-0.2em}{\sigma}_{y}^{p}\sin(2\Omega_{1}t)]\nonumber.
\end{align}
Following the application of the approximation $4E_{x}\gg g$ and $2\omega_{ac}\gg g$, the resonance condition $2E_{x}-2\Omega_{1}-2\Omega_{2}=\omega_{ac}$ is validated, leading to the sensing of AC field signal (as will be demonstrated below).\par

\begin{figure}
  \centering
\includegraphics[width=3.3 in]{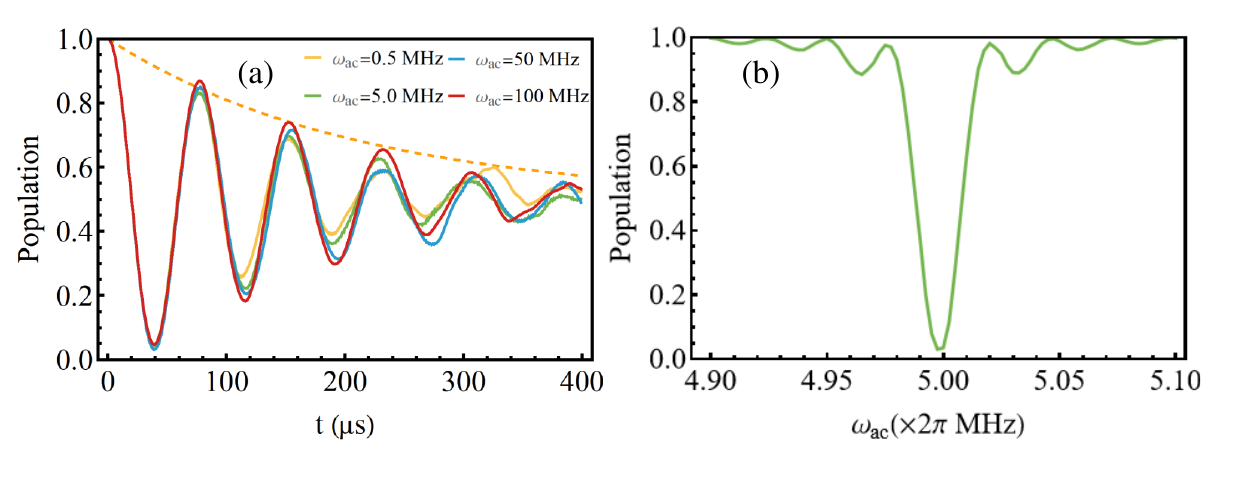}
  \caption{(a) Simulations of population of solid-state spin of $|0\rangle$ state, when there is a target AC field with different frequency $\omega_{ac}$. The expectation coherence time $T_{2}^{\ast}=0.1$ $\mu s$. Assuming $\Omega_{1}\gg\Omega_{2}$, we choose $E_{x}=(2\pi)\times110$ MHz, $g=(2\pi)\times0.1$ MHz and make $\omega_{ac}/(2\pi)=0.5$(orange), $5.0$(green), $50$(blue), $100$(red) MHz to satisfy the resonance condition $2E_{x}-2\Omega_{1}-2\Omega_{2}=\omega_{ac}$. The corresponding  $\Omega_{1}/(2\pi)$ are $99.77$, $97.73$, $77.27$, $54.55$ MHz and $\Omega_{2}/(2\pi)$ are $9.98$, $9.77$, $7.73$, $5.45$ MHz. The results are averaged for $100$ runs and a relative amplitude error is $0.005$. (b) The detected spectrum of AC field with frequency $(2\pi)\times$5.0 MHz and evolution time $t=40$ $\mu s$. Other parameters are the same as the green line in (a). }\label{protocol12}
\end{figure}

\begin{figure}[h]
  \centering
  \includegraphics[scale=0.43]{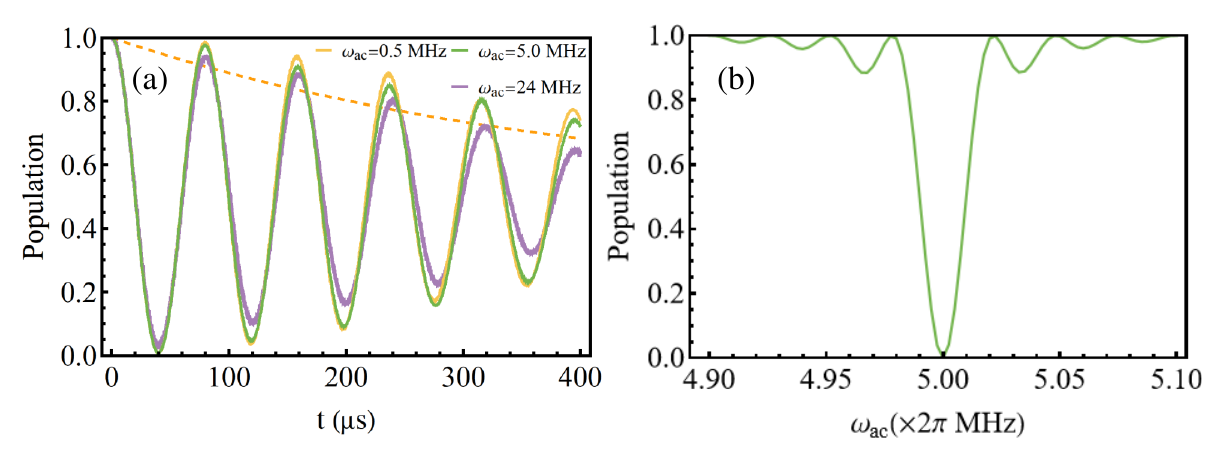}
  \caption{(a) The AC field signal with different $\omega_{ac}$ under reduced ZFS. The expectation coherence time $T_{2}^{\ast}=0.3$ $\mu s$. Assuming $\Omega_{1}=10\Omega_{2}$, we choose $E_{x}=(2\pi)\times24$ MHz, $g=(2\pi)\times0.1$ MHz and make $\omega_{ac}/(2\pi)=0.5$(orange); $5.0$(green); $24$(purple) MHz to satisfy the resonance condition. The corresponding  $\Omega_{1}/(2\pi)$ are $21.59$, $19.55$, $10.91$ MHz and $\Omega_{2}/(2\pi)$ are $2.16$, $1.95$, $1.09$ MHz. The results are averaged for $100$ runs and a relative amplitude error is $0.005$. (b) The detected spectrum of AC field with frequency $(2\pi)\times$5.0 MHz and evolution time $t=40$ $\mu s$. Other parameters are the same as the green line in (a). }\label{protocol12}
\end{figure}

Leveraging the robustness against the system noise $\delta E$ and amplitude noise $\delta\Omega_{1}(t)$, we utilize our phase modulation-optimized control scheme to probe AC fields of varying frequencies. The effective Hamiltonian of AC fields sensing for numerical simulation is described in Eq.(35). As a first example, we consider an electron spin in $Cr((trimethylsilyl)methyl)_{4}$ molecular sample, which exhibits a transverse ZFS of 110 MHz \cite{laorenza2021tunable}.
Because of the controllable Rabi frequency of the MW driving, our control method enables the detection of AC fields in the range $(2\pi)\times[0.5,100.0]$ MHz with $E_{x}=(2\pi)\times110$ MHz. Notably, the coherence time $T_{2}$ of the system is extended from 0.1 $\mu s$ to several hundred microseconds (see Fig. 5(a)). Figure 5(a) shows a simulation of the population evolution of the  $|0\rangle$ state (represented as $|\mu_{-}\rangle_{p}$ in our scheme) for sensing different AC fields. The improvement in coherence time by three orders of magnitude is primarily ascribed to the robustness of our method against system noise. Moreover, under strong transverse zero-field splitting, hyperfine interactions are quadratically suppressed, rendering their influence negligible. The remaining decay might be due to the second order noise term of the first drive $\sim\delta\Omega_{1}(t)^{2}/\Omega_{2}$. Here comes another example of a shallow NV center in nanodiamond, where $E_{x}=(2\pi)\times24$ MHz \cite{awadallah2023spin}, we demonstrate that our method retains effective for sensing a wide range of AC fields, from several hundred kHz to tens of MHz (see Fig. 6(a)).  The dephasing time $T_{2}$ still attains hundreds of microseconds, owing to the enhanced robustness against amplitude noise with reduced $E_{x}$. The spectrum of the sensed AC field with $\omega_{ac}=(2\pi)\times5.0$ MHz is shown in Fig. 6(b).

The sensitivity is improved due to two distinct factors: our control method enables the color center to respond linearly to the AC magnetic field and prolongs the coherence time, thereby allowing AC field sensing. The improvement could be estimated as $g\sqrt{T_{2}^{\ast}}/E_{x}\sqrt{T_{2}}$. However, our method is limited by the resonance condition $2E_{x}-2\Omega_{1}-2\Omega_{2}=\omega_{ac}$, where $\Omega_{1}\gg\Omega_{2}$ is required. Therefore, the largest detectable frequency of the AC field is in the order of the transverse ZFS. Moreover, in the lab, there is also a limitation of the amplitude of MW Rabi frequency, if we have a sensor with $E_{x}\sim$ GHz, the detectable frequency will be close to $2E_{x}$ and it is not possible to detect an AC field whose frequency is smaller than 1 GHz. Despite these limitations, our method presents significant advantages over conventional linear MW driving schemes. Compared to molecular clock sensors with dynamical decoupling control \cite{JiangChen2024}, when the transverse ZFS of a color center is given, the detectable frequency range of our scheme can be much larger. For example, with a transverse ZFS of 110 MHz, we have shown a broader band for AC field detection, spanning from hundreds of kHz to hundreds of MHz. Consequently, our method enables wide band sensing of AC fields, facilitated by the extensive range of transverse ZFS available in solid-state spin systems, such as shallow NV centers in nanodiamonds \cite{awadallah2023spin} with more than 10 MHz, those in SiC \cite{luo2023fabrication} and molecular color centers \cite{laorenza2021tunable} with over 100 MHz, along with our MW control scheme and adjustable Rabi frequency.\par

\vspace{-1em}
\section{Conclusion}
 In solid-state color centers, clock transitions induced by transverse ZFS effectively reduce sensitivity to first-order magnetic field noise, thereby enhancing the coherence time of the system. However, these color centers cannot be exploited for magnetometric applications due to their insensitivity to magnetic signals. Our proposed control scheme addresses this challenge and enables the utilization of the solid-state color centers for wide band AC field detection. Additionally, our control scheme demonstrates robustness against the system noise, with phase modulation effectively suppressing first-order amplitude noise. Furthermore, our approach significantly improves the coherence time of the system, enhancing sensitivity for magnetometric applications and opening new avenues for the deployment of solid-state spin systems in a variety of sensing technologies.\par

\vspace{+1em}
\section*{Acknowledgements}
We gratefully acknowledge the discussions with Martin. B. Plenio. X.Y. Ding, L. Chen and Q. Chen are supported by Natural Science Foundation of China (Grants No. 12247105,12375012), Hunan provincial major scitech program (Grants No. 2023zk1010, XJ2302001).

Xiayan Ding and Lu Chen contributed equally to this work.

 \bibliographystyle{unsrt}

\end{document}